\documentclass{article}

    \usepackage[preprint, nonatbib]{neurips_2023}

\usepackage[utf8]{inputenc} 
\usepackage[T1]{fontenc}    
\usepackage{hyperref}       
\usepackage{url}            
\usepackage{booktabs}       
\usepackage{amsfonts}       
\usepackage{nicefrac}       
\usepackage{microtype}      
\usepackage{xcolor}         
\usepackage{graphicx}
\usepackage{enumitem}
\usepackage{tabularx}
\usepackage{array}
\usepackage{enumitem}
\usepackage{appendix}  
\usepackage[section]{placeins}
\setlist[enumerate,1]{leftmargin=20pt,topsep=0pt,itemsep=0pt,partopsep=0pt, parsep=0pt}

\title{Second Sight: Using brain-optimized encoding models to align image distributions with human brain activity}

\author{Reese Kneeland \\
Department of Computer Science\\
University of Minnesota\\
Minneapolis MN, 55455 \\
\texttt{rek@umn.edu} \\
\And
Jordyn Ojeda \\
Department of Computer Science\\
University of Minnesota\\
Minneapolis MN, 55455 \\
\texttt{ojeda040@umn.edu} \\
\And
Ghislain St-Yves \\
Department of Neuroscience\\
University of Minnesota\\
Minneapolis MN, 55455 \\
\texttt{gstyves@umn.edu} \\
\And
Thomas Naselaris \\
Department of Neuroscience\\
University of Minnesota\\
Minneapolis MN, 55455 \\
\texttt{nase0005@umn.edu} \\
}

\begin{document}{

\maketitle

\begin{abstract}
Two recent developments have accelerated progress in image reconstruction from human brain activity: large datasets that offer samples of brain activity in response to many thousands of natural scenes, and the open-sourcing of powerful stochastic image-generators that accept both low- and high-level guidance. Most work in this space has focused on obtaining point estimates of the target image, with the ultimate goal of approximating literal pixel-wise reconstructions of target images from the brain activity patterns they evoke. This emphasis belies the fact that there is always a family of images that are equally compatible with any evoked brain activity pattern, and the fact that many image-generators are inherently stochastic and do not by themselves offer a method for selecting the single best reconstruction from among the samples they generate. We introduce a novel reconstruction procedure (Second Sight) that iteratively refines an image distribution to explicitly maximize the alignment between the predictions of a voxel-wise encoding model and the brain activity patterns evoked by any target image. We use an ensemble of brain-optimized deep neural networks trained on the Natural Scenes Dataset (NSD) as our encoding model, and a latent diffusion model as our image generator. At each iteration, we generate a small library of images and select those that best approximate the measured brain activity when passed through our encoding model. We extract semantic and structural guidance from the selected images, used for generating the next library. We show that this process converges on a distribution of high-quality reconstructions by refining both semantic content and low-level image details across iterations. Images sampled from these converged image distributions are competitive with state-of-the-art reconstruction algorithms. Interestingly, the time-to-convergence varies systematically across visual cortex, with earlier visual areas generally taking longer and converging on narrower image distributions, relative to higher-level brain areas. Second Sight thus offers a succinct and novel method for exploring the diversity of representations across visual brain areas.
\end{abstract}

\section{Introduction}

Reconstructing visual stimuli from measured brain activity is a long-standing problem in computational neuroscience. Improvements bring us closer to understanding visual representation in the brain by providing interpretable tools for examining brain states \cite{naselaris2011, St-Yves_heirarchy}. 

Two recent developments have accelerated progress in image reconstruction from human brain activity: large datasets that offer samples of brain activity in response to many thousands of natural scenes \cite{Allen2021a}, and the open-sourcing of powerful image-generators that can invert abstract feature representations into distributions over natural scenes \cite{stablediffusion}. In many recent approaches, these data are used to map brain activity directly to feature representations that are fed into the generator.

Although this approach has been proven quite successful \cite{lin2022mind,ozcelik2023braindiffuser,gu2023decoding, Takagi2022.11.18.517004, lu2023minddiffuser, St-Yves_gan}, it has clear limitations. First, latent representations that are used to drive image generators may overlap with only a subset of the rich and diverse representations encoded in brain activity across visual cortex \cite{naselaris2011}.

Second, even with the very large datasets that have recently been distributed, bottom-up approaches that seek to decode only low-level representations do not yield semantically meaningful reconstructions \cite{NASELARIS2009902}. Thus, most successful pipelines extract some form of semantic representation in conjunction with low-level structural representations to guide reconstruction. This practice requires a policy for weighting the relative influence of high-level semantic representations and low-level structural representations on the reconstruction. Semantic representations are structurally ambiguous, so weighting them strongly will result in broad output distributions that capture semantics at the expense of uncertainty about structure; strongly weighting structural representations offers lower structural uncertainty at the expense of semantic accuracy. Because of this tension, some amount of stochasticity in the reconstructions may be inevitable \cite{St-Yves_gan}. The problem of how to reduce structural uncertainty without distorting semantic content is currently unsolved.

Finally, given that some amount of stochasticity in the reconstruction pipeline must be accepted, a method for scoring individual candidate reconstruction sampled from a distribution over images is also desirable.

We develop an approach to reconstructing natural images that addresses all three problems. Rather than training decoding models that minimize loss in the space of high- or low-level representations, we train “brain-optimized” encoding models that minimize loss in the space of brain activity. At test, we use the encoding model to select image distributions that are closely aligned to the brain activity pattern evoked by the target image. Specifically, we select images from small libraries generated by a diffusion model \cite{stablediffusion}. By using encoding models optimized to predict activity in most of the visual cortex, we leverage the rich diversity of representations in the human visual system. By incrementally lowering the breadth of reconstructed image distributions, and guiding it with selections from sequential search iterations, we are able to minimize uncertainty about structural detail while preserving and even improving the semantic aspects of the reconstruction. In this way, we are able to leverage the stochasticity inherent in diffusion models, progressively reducing the width of the generated image distribution until it aligns with the ambiguity measured in cortical representations of natural stimuli. We refer to our pipeline as ``Second Sight".

We show that selecting the most brain-aligned image distribution from each search yields semantically meaningful reconstructions that often correctly reflect low-level details of target images (e.g., pose). Our reconstructions are competitive with SOTA systems in terms of their proximity to ground truth images in a variety of feature spaces and, as expected, establish a new SOTA in terms of their alignment to activity patterns in the brain.

Furthermore, by analyzing search dynamics we show that, for most brain areas, samples from output distributions exceed ground truth alignment after several iterations, demonstrating that any single pattern of brain activity admits a distribution of highly aligned images. We show how this degeneracy increases across visual cortex, as evidenced by the rapid rate at which reconstructions align to activity in high-level areas, relative to low-level areas. We argue that “rate of convergence” is a succinct and attractive metric of representational invariance across brain areas and brain states.

\section{Approach}

\begin{figure}[!htb]
\begin{center}
\includegraphics[width=\columnwidth]{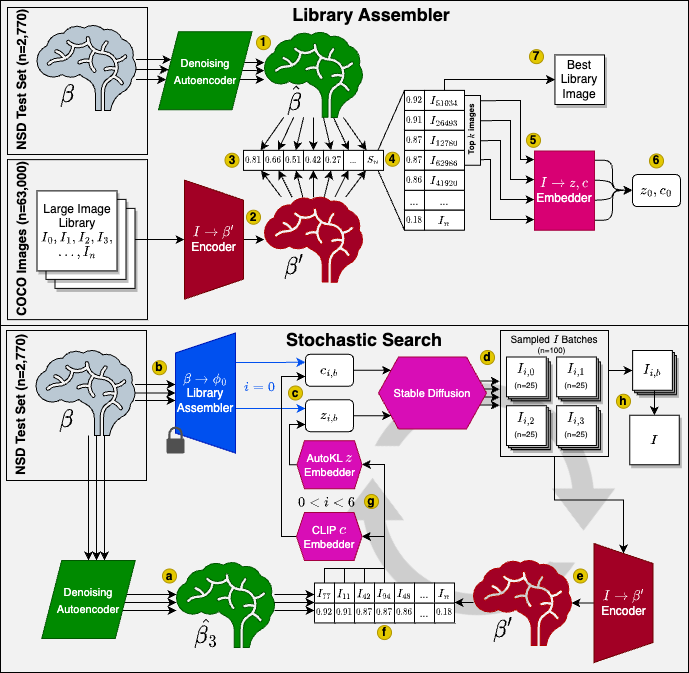}
\end{center}
\caption{Second Sight pipeline diagram. The pipeline consists of two stages. In the first stage (Library Assembler) we assemble the feature vectors $z_0$ and $c_0$ that seed the second stage (Stochastic Search), in which we iteratively align an image distribution to brain activity. Details on the numbered and lettered components of each stage are detailed in Sections \ref{library} and \ref{search}.} 
\label{figure:pipeline}
\end{figure}

\subsection{Dataset}
We use the Natural Scenes Dataset (NSD) \cite{Allen2021a}, which comprises 27,749 scans distributed across the 37 sessions publicly released by NSD for subjects 1, 2, 5, and 7, which are the subjects that completed all scanning sessions. We partition the data for each subject into three distinct sets: training (n=20,809), validation (n=4,170), and test (n=2,770). The test set is derived from a shared pool of 1,000 images that all subjects have seen, whereas the training and validation sets are generated from the remaining data. We withhold 85 test set images that either overlap the Stable Diffusion training dataset \cite{schuhmann2022laion5b} or were used for manual qualitative tuning of various hyperparameters in our method. For further details on what data was withheld see Appendix A. In this investigation, we mask preprocessed fMRI signals using the nsdgeneral mask in 1.8mm resolution. The ROI consists of [15724,14278,13039,12682] voxels for the four subjects and includes the entire visual cortex region. For further information regarding data preprocessing, please refer to \cite{Allen2021a}. In some cases, we further segment the nsdgeneral ROI into V1, V2, V3, V4 (collectively called early visual cortex) and higher visual areas (the set complement of nsdgeneral and early visual cortex).

\subsection{Models}
The pipeline's model preparation phase comprises several integral components, some of which are pre-trained implementations and some we prepare for this work. For the Library Assembler, we load a pre-trained OpenClip ViT-H/14 \cite{radford2021learning, openclip, openclip2} image embedding model for the $I \rightarrow c$ step, and a pre-trained VDVAE decoder \cite{vdvae} for the $I \rightarrow z$ step. The ensemble of encoding models includes a brain-optimized neural network (GNet; \cite{St-Yves_heirarchy}), and an independent 3-layer MLP that maps the ViT-H/14 CLIP image embedding of an input image to a predicted brain activity pattern that includes all of nsdgeneral. Both encoding models were trained on the NSD training/validation sets. Unless noted otherwise, we averaged the outputs of this ensemble of encoding models to obtain a single predicted activity pattern $\beta'$. We refer to this average as the output of a ``hybrid” encoding model. While more nuanced methods for combining encoding model outputs exist \cite{Lin2023.04.23.537988}, this methodology produces an encoding model with sufficient sensitivity to both structural and semantic features, more details of which can be found in Appendix B.5. The encoded $\beta'$s corresponding to the ground truth images serve as targets in conjunction with the original fMRI scans to train a denoising autoencoder. This process enables more robust comparisons between the original brain data and the encoded predictions by removing some of the noise associated with fMRI data. This work uses the Stable Diffusion~\cite{stablediffusion} Reimagine model (SD) as our image generator. The SD model is adapted in our implementation to receive a latent $z$ vector and a strength parameter in addition to a CLIP embedding. More details of the individual model components and our pipeline can be found in Appendix B.

\subsection{Stage 1: Library Assembler}
\label{library}

Numbered items reference Figure \ref{figure:pipeline}, (top). 

\begin{enumerate}
\item Three samples of measured brain activity $\beta$ for a target image in the test set are passed through an autoencoder and concatenated to yield denoised activity patterns $\hat{\beta}$. 
\item Each of the $n=63,000$ images ($I$) in a large image library  is passed through an encoding model (either a hybrid model or the GNet model), yielding a predicted activity pattern $\beta'$ for each library image. The library is a subset of the COCO dataset \cite{microsoftcoco}. None of the library images were included in our training, test, or validation sets.

\item Each image in the large library is scored by calculating Pearson correlation between $\hat{\beta}$ and $\beta'$. We refer to this form of scoring as ``brain correlation”. Where noted, brain correlation scores may also be computed with respect to activity in a single ROI (e.g., V1).  
\item Library images are sorted in decreasing order of their brain correlation scores, and the top $k$ images are selected.
\item The top $k$ images are embedded into either a high-level or a low-level representation. The high-level feature representation is a ViT-H/14 CLIP vector $c$, with $k=100$. The low-level representation is the first 31 components of the VDVAE latent vector \cite{vdvae}, with $k$ parameter $=25$. For high-level representation, the top $k$ selections are made using brain correlation between the hybrid encoding model and all voxels in nsdgeneral. For low-level representations the top $k$ selections are made using brain correlation between the GNet model and all voxels in early visual cortex. The VDVAE representations are converted to pixels using the pre-trained VDVAE decoder and then passed through SD's AutoKL embedder to generate the low-level structural guidance $z$.

\item The top $k$ high-level (low-level) representations are averaged to yield seed representations $c_0$ ($z_0$). 
\item The top-scoring image from the library is also saved as an alternative encoding based reconstruction approach. We refer to this method as the ``Best Library Image".
\end{enumerate}

\subsection{Stage 2: Stochastic Search}
\label{search}

Lettered items refer to Figure \ref{figure:pipeline} (bottom).

\begin{enumerate}[label=(\alph*)]
\item As above, three samples of brain activity patterns evoked by the test target image are denoised and concatenated ($\hat{\beta}$).
\item Representations $c_0$ and $z_0$ (produced by the Library Assembler) seed the first search iteration (blue).
\item High-level ($c_0$) and low-level ($z_0$) representations are input to the image generator. Iterations $i > 0$ yield multiple high-level and low-level representations, denoted $c_{i,b}$ and $z_{i,b}$, $0 < i < 6 $, $0 < b < 3$.

\item Image batches ($I_{i, b}$) are sampled from SD conditioned on $c_{i,b}$ and $z_{i, b}$. SD exposes a strength parameter $\lambda$ that modulates the quantity of noise introduced to the latent $z$ vector, with a value of 1 representing the addition of 100\% noise and 0 signifying the addition of 0\% noise. Furthermore, the strength parameter dictates the number of CLIP-guided denoising steps to be executed, as $z$ vectors with reduced noise necessitate less denoising. This parameter effectively controls the breadth of the distribution of generated images originating from a given $z$ vector.

\item  Images in each batch are passed through the encoding model to yield predicted brain activity $\beta’$.
\item Images are scored and sorted as in Stage 1.  
\item The four images with the best correlation scores are embedded into their latent $c$ and $z$ vectors. By using the top four images to seed parts of the following iteration, we ensure our search does not get stuck in a local maximum. The $z$ vectors from each of the four batch seeds $b$ act as the corresponding $z_{i,b}$ vector for the $n_b$ images in the forthcoming batch $I_{i,b}$. We observed that using the embedded $c$ vector directly in the next iteration led to semantic instability, which we correct by spherically interpolating between the current set of $c$ vectors and the previous iteration's best CLIP vector $c_{i-1}$ according to a momentum parameter $\gamma$. This produces our four $c_{i,b}$ vectors, each mixing a little of the semantics from the new batch seeds $b$ into the next iteration of batches $I_{i,b}$. $n_b$, $\gamma$, and the strength parameter $\lambda$ are progressively reduced across iterations in accordance with monotonically decreasing exponential or cubic schedules (described in Appendix B.7. )
\item The image distribution $I_{i,b}$ that yields the highest average brain correlation throughout all iterations is selected as the output distribution, from which the image with the highest brain correlation $I$ is selected as the output image displayed in Figures \ref{figure:subjectcomp} and \ref{figure:papercomp}.
\end{enumerate}

\begin{figure}[!ht]
\begin{center}
\includegraphics[width=\columnwidth]{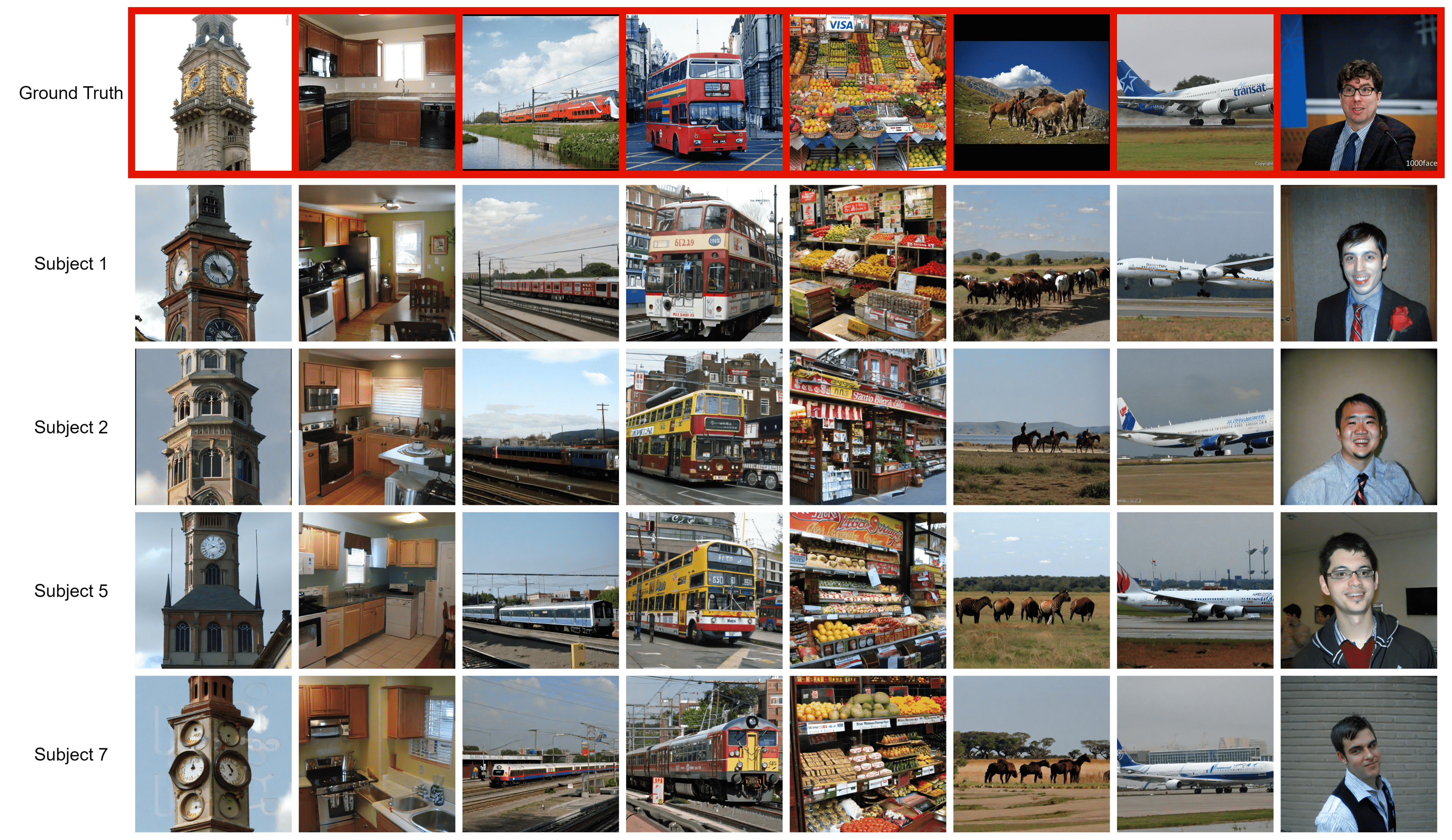}
\end{center}
\caption{Example reconstructions. The first row is the ground truth image, indicated by the red background, while the remaining rows are the reconstructions for each individual subject (1, 2, 5, and 7). Some additional samples can be found in Appendix E.} 
\label{figure:subjectcomp}
\end{figure}

\section{Results}

\begin{figure}[!ht]
\begin{center}
\includegraphics[width=\columnwidth]{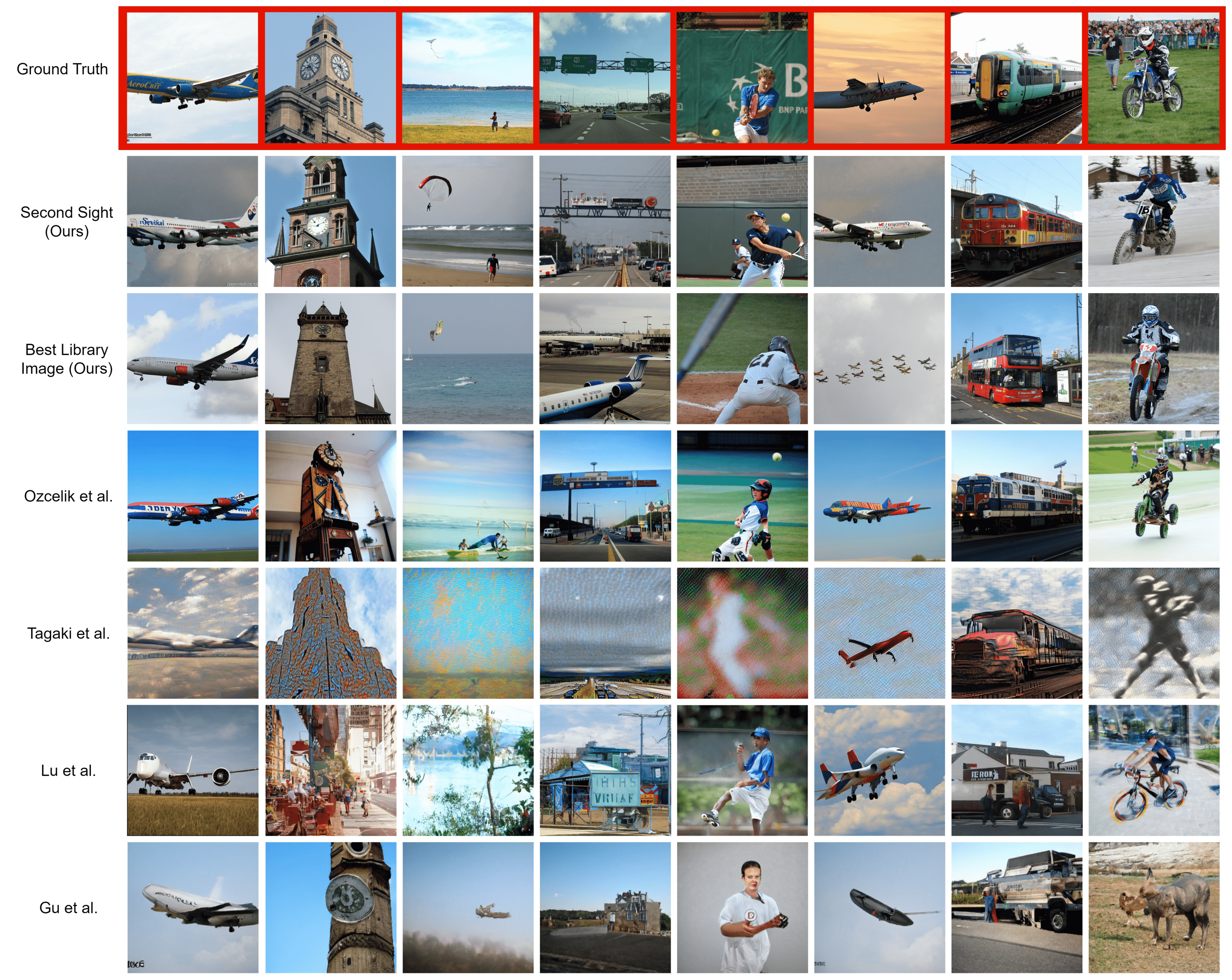}
\end{center}
\caption{Comparative assessment of reconstructions. The first row is the ground truth image, indicated by the red background, the second row is Second Sight reconstructions, the third row shows the "Best Library Image" approach, while the remaining rows represent subject 1 results from previous works in Ozcelik et al. \cite{ozcelik2023braindiffuser}, Tagaki et al. \cite{Takagi2022.11.18.517004}, Lu et al. \cite{lu2023minddiffuser}, and Gu et al. \cite{gu2023decoding}} 
\label{figure:papercomp}
\end{figure}

\begin{table}[h]
\centering
\resizebox{\columnwidth}{!}{%
\begin{tabular}{lccccc}
\hline
\multicolumn{1}{|c|}{Methods}                   & \multicolumn{5}{c|}{Low Level Structural Metrics}                                                                                                                                              \\ \cline{2-6} 
\multicolumn{1}{|c|}{}                          & \multicolumn{1}{c|}{PixCorr↑}        & \multicolumn{1}{c|}{SSIM↑}         & \multicolumn{1}{c|}{AlexNet(2)↑}     & \multicolumn{1}{c|}{AlexNet(5)↑}     & \multicolumn{1}{c|}{AlexNet(7)↑}     \\ \hline
\multicolumn{1}{|l|}{Second Sight (ours)}       & \multicolumn{1}{c|}{.185}            & \multicolumn{1}{c|}{.288}          & \multicolumn{1}{c|}{91.7\%}          & \multicolumn{1}{c|}{95.0\%}          & \multicolumn{1}{c|}{94.2\%}          \\
\multicolumn{1}{|l|}{Best Library Image (ours)} & \multicolumn{1}{c|}{.138}               & \multicolumn{1}{c|}{.255}             & \multicolumn{1}{c|}{86.9\%}               & \multicolumn{1}{c|}{93.1\%}               & \multicolumn{1}{c|}{93.3\%}               \\
\multicolumn{1}{|l|}{Ozcelik et al. \cite{ozcelik2023braindiffuser}}            & \multicolumn{1}{c|}{\textbf{.304}}   & \multicolumn{1}{c|}{\textbf{.294}} & \multicolumn{1}{c|}{\textbf{96.7\%}} & \multicolumn{1}{c|}{\textbf{97.6\%}} & \multicolumn{1}{c|}{\textbf{96.3\%}} \\
\multicolumn{1}{|l|}{Tagaki et al.\cite{Takagi2022.11.18.517004}}             & \multicolumn{1}{c|}{.269}            & \multicolumn{1}{c|}{.217}          & \multicolumn{1}{c|}{75.1\%}          & \multicolumn{1}{c|}{77.0\%}          & \multicolumn{1}{c|}{76.0\%}          \\
\multicolumn{1}{|l|}{Lu et al.\cite{lu2023minddiffuser}}                 & \multicolumn{1}{c|}{.246}            & \multicolumn{1}{c|}{.221}          & \multicolumn{1}{c|}{74.7\%}          & \multicolumn{1}{c|}{78.9\%}          & \multicolumn{1}{c|}{77.1\%}          \\
\multicolumn{1}{|l|}{Gu et al.\cite{gu2023decoding}}   & \multicolumn{1}{c|}{.146}  & \multicolumn{1}{c|}{.279}   & \multicolumn{1}{c|}{79.4\%}  & \multicolumn{1}{c|}{90.1\%} & \multicolumn{1}{c|}{93.7\%}               \\ \hline
                                                & \multicolumn{1}{l}{}                 & \multicolumn{1}{l}{}               & \multicolumn{1}{l}{}                 & \multicolumn{1}{l}{}                 & \multicolumn{1}{l}{}                 \\ \hline
\multicolumn{1}{|c|}{Methods}                   & \multicolumn{5}{c|}{High Level Semantics Metrics}                                                                                                                                              \\ \cline{2-6} 
\multicolumn{1}{|c|}{}                          & \multicolumn{1}{c|}{CLIP(2-way)↑}    & \multicolumn{1}{c|}{CLIP(cos)↑}    & \multicolumn{1}{c|}{Inception V3↑}   & \multicolumn{1}{c|}{EffNet-B↓}       & \multicolumn{1}{c|}{SwAV↓}           \\ \hline
\multicolumn{1}{|l|}{Second Sight (ours)}       & \multicolumn{1}{c|}{88.3\%}          & \multicolumn{1}{c|}{\textbf{.664}} & \multicolumn{1}{c|}{84.7\%}          & \multicolumn{1}{c|}{.776}            & \multicolumn{1}{c|}{.418}            \\
\multicolumn{1}{|l|}{Best Library Image (ours)} & \multicolumn{1}{c|}{85.9\%}               & \multicolumn{1}{c|}{.657}             & \multicolumn{1}{c|}{82.5\%}               & \multicolumn{1}{c|}{.798}               & \multicolumn{1}{c|}{.445}               \\
\multicolumn{1}{|l|}{Ozcelik et al\cite{ozcelik2023braindiffuser}}           & \multicolumn{1}{c|}{\textbf{92.6\%}} & \multicolumn{1}{c|}{.657}          & \multicolumn{1}{c|}{\textbf{88.5\%}} & \multicolumn{1}{c|}{\textbf{.764}}   & \multicolumn{1}{c|}{\textbf{.411}}   \\
\multicolumn{1}{|l|}{Tagaki et al.\cite{Takagi2022.11.18.517004}}             & \multicolumn{1}{c|}{64.4\%}          & \multicolumn{1}{c|}{.566}          & \multicolumn{1}{c|}{64.0\%}          & \multicolumn{1}{c|}{.954}           & \multicolumn{1}{c|}{.668}           \\
\multicolumn{1}{|l|}{Lu et al.\cite{lu2023minddiffuser}}                 & \multicolumn{1}{c|}{70.8\%}          & \multicolumn{1}{c|}{.578}          & \multicolumn{1}{c|}{69.6\%}          & \multicolumn{1}{c|}{.913}           & \multicolumn{1}{c|}{.552}           \\
\multicolumn{1}{|l|}{Gu et al.\cite{gu2023decoding}}  & \multicolumn{1}{c|}{78.5\%}  & \multicolumn{1}{c|}{.597} & \multicolumn{1}{c|}{79.0\%}  & \multicolumn{1}{c|}{.864}  & \multicolumn{1}{c|}{.451}       \\ \hline
\end{tabular}%
}
\caption{Quantitative comparison against past reconstruction methods on subject 1. For each measure, the best value is in bold. For EffNet-B and SwAV distances, lower is better. Higher is better for all other metrics. This is indicated by the arrow pointing up or down.}
\label{table:results}
\end{table}

\subsection{Assessment of reconstruction quality}

Figure \ref{figure:subjectcomp} shows example reconstructions for four subjects. As shown, in the best case reconstructions recapitulate both the semantic content of images and their low-level structural details for all subjects. Figure \ref{figure:papercomp} presents a comparison of reconstructions from a number of recent works. At a glance, our reconstructions compare favorably.

To make explicit comparisons of reconstruction quality across different approaches, we calculated the proximity of our and others' reconstructions to ground truth in a variety of feature spaces (Table \ref{table:results}) and the space of brain activity (Table \ref{table:braincorr}). To calculate these metrics, we contacted the authors of all of the previous works we compare against to obtain the full set of their generated results. Wherever possible, we calculated the metrics as an average of 5 generated images sampled from the final image distribution of the method. The numbers for Second Sight are calculated on 5 images sampled from the distribution with the highest average brain correlation with the nsdgeneral mask of the visual cortex. The numbers for our "Best Library Image" were calculated on a single image per sample, as this method is deterministic and does not produce a distribution. Although we performed no explicit optimization in feature space, our reconstructions are competitive with current SOTA at approximating ground truth features. As expected, our reconstructions establish SOTA brain correlation scores for all visual ROIs we examined. For more details on how metrics were calculated as well as results in a multi-subject analysis, see Appendix C.

\begin{table}[h]
\centering
\resizebox{\columnwidth}{!}{%
\begin{tabular}{|l|cccccc|}
\hline
\multicolumn{1}{|c|}{Methods} & \multicolumn{6}{c|}{Brain Correlation Metrics}                                                                                                        \\ \cline{2-7} 
\multicolumn{1}{|c|}{}        & \multicolumn{1}{c|}{V1↑} & \multicolumn{1}{c|}{V2↑} & \multicolumn{1}{c|}{V3↑} & \multicolumn{1}{c|}{V4↑} & \multicolumn{1}{c|}{Higher Vis↑} & nsdgeneral↑ \\ \hline
Second Sight (ours)           & \multicolumn{1}{c|}{\textbf{.739}}  & \multicolumn{1}{c|}{\textbf{.774}}  & \multicolumn{1}{c|}{\textbf{.817}}  & \multicolumn{1}{c|}{\textbf{.854}}  & \multicolumn{1}{c|}{\textbf{.932}}          & \textbf{.905}           \\
Best Library Image (ours)     & \multicolumn{1}{c|}{.697}  & \multicolumn{1}{c|}{.750}  & \multicolumn{1}{c|}{.800}  & \multicolumn{1}{c|}{.847}  & \multicolumn{1}{c|}{.927}          & .895           \\
Ozcelik et al.\cite{ozcelik2023braindiffuser}  & \multicolumn{1}{c|}{.672}  & \multicolumn{1}{c|}{.676}  & \multicolumn{1}{c|}{.705}  & \multicolumn{1}{c|}{.757}  & \multicolumn{1}{c|}{.862}   & .828  \\
Tagaki et al.\cite{Takagi2022.11.18.517004}   & \multicolumn{1}{c|}{.324}  & \multicolumn{1}{c|}{.271}  & \multicolumn{1}{c|}{.282}  & \multicolumn{1}{c|}{.307}  & \multicolumn{1}{c|}{.361}    & .356     \\
Lu et al.\cite{lu2023minddiffuser}   & \multicolumn{1}{c|}{.286}  & \multicolumn{1}{c|}{.276}  & \multicolumn{1}{c|}{.304}  & \multicolumn{1}{c|}{.364}  & \multicolumn{1}{c|}{.511}    & .467           \\
Gu et al.\cite{gu2023decoding}   & \multicolumn{1}{c|}{.397}  & \multicolumn{1}{c|}{.425}  & \multicolumn{1}{c|}{.460}  & \multicolumn{1}{c|}{.550}  & \multicolumn{1}{c|}{.740}  & .676           \\ \hline
\end{tabular}%
}
\caption{Evaluation of brain correlation scores against past reconstruction methods on subject 1. This metric evaluates how well the reconstructed images represent activity in various brain regions according to our hybrid encoding model.}
\label{table:braincorr}
\end{table}

\subsection{Analysis of search dynamics}

Figure \ref{figure:refinement} shows examples of the best-aligned reconstructions across iterations of several searches. As seen in these examples, using an encoding model to iteratively align reconstructions to brain activity results in gradual refinement of both the semantic content (i.e., the category of objects depicted) and structural details (i.e., their pose and composition) of the reconstructions. For more details on this ablation study, see Appendix D.

\begin{figure}[h]
\begin{center}
\includegraphics[width=\columnwidth]{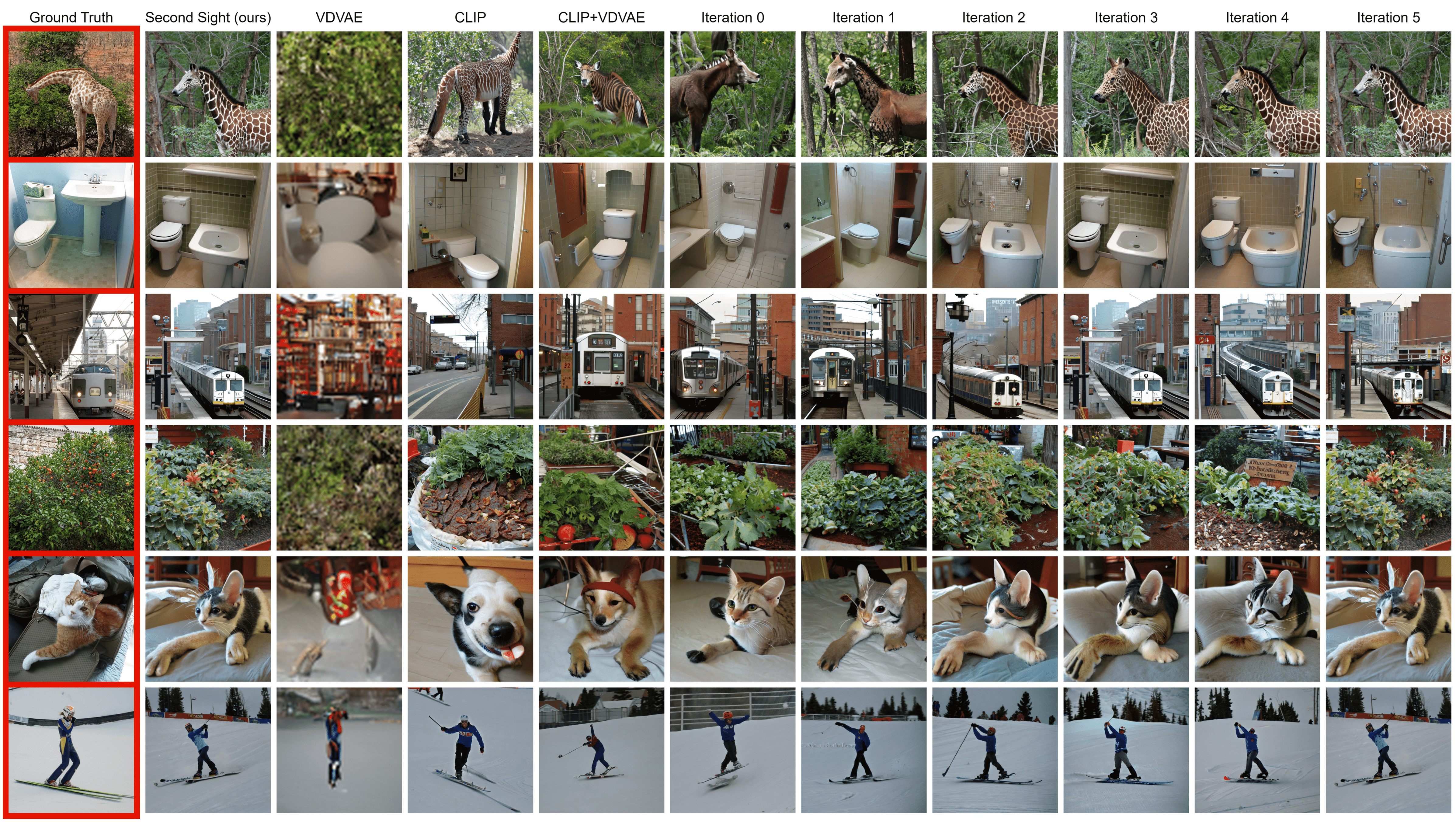}
\end{center}
\caption{Examples of best reconstructions across iterations of the search process. ``VDVAE" indicates reconstructions generated from a VDVAE with low-level input prepared by the Library Assembler. ``CLIP" indicates reconstructions generated by SD using CLIP guidance ($c_0$) prepared by the Library Assembler. ``CLIP+VDVAE" indicates reconstructions generated by SD using CLIP guidance ($c_0$) and structural guidance ($z_0$) prepared by the Library Assembler.} 
\label{figure:refinement}
\end{figure}

Given the degeneracy of representations encoded in brain activity, any search through image space can in principle identify images that, when passed through our encoding model, yield predictions that are more highly correlated with brain activity than the target image that evoked it. We analyzed the evolution of the alignment between brain activity and our reconstructions to determine at what point the process exceeded the alignment of the ground truth target images (Figure~\ref{figure:plot}, plots A and B). In this analysis, surpassing the brain correlation score of the ground truth image serves as functional confirmation that we have converged on a distribution of images that aligns closely with the brain's representation of the original stimuli.

In high-level areas, alignment to brain activity achieves parity with ground truth by iteration 0 (on average), while in V1 average alignment is still below ground truth by iteration 5 (the highest iteration tested here). Interestingly, the "rate of convergence" to parity with ground truth orders brain areas by their level of representational invariance, with V1 converging most slowly, followed by V2, V3, V4, and higher-level visual cortex. 

We also determined the relative increase in alignment over several ``unrefined" reconstructions sampled at random from output distributions guided only by low-level representations (``VDVAE"), only by high-level representations (``CLIP"), or both (``CLIP+VDVAE"). In V1, the percent of reconstructions that achieve parity with ground truth is quite high for the "low-level" only reconstructions but is eventually exceeded at iteration 6 (see Figure \ref{figure:refinement} for visualizations). Analysis of the subset of reconstructions that achieved parity with ground truth for all brain areas (Figure \ref{figure:plot} plot C) revealed a similar ordering of brain areas by their rate of convergence to parity with ground truth.

\section{Limitations}

One limitation of the current implementation is the rather small size of the fixed Large Library used in the Library Assembler and the smaller libraries generated throughout the search. Small library sizes make it difficult to break correlations between object categories. For example, trains are prevalent in the COCO dataset used as the large library for this work, and are often depicted near buildings. If brain activity encodes a picture of a building, it is likely that a picture of a train will be included in the subset of images with the highest brain correlation scores. Of course, this limitation becomes an advantage when reconstructing images that depict trains, but is indicative of larger biases in the prior enforced by the image library that might constrain this method's ability to represent smaller image categories or finer semantic details within images. The prior is also entirely restricted to the image content and classes present within the library, as object classes that are out of distribution for the large library would never be represented in the reconstructions.

For the technical limitations of our implementation, the most immediate limitation lies in our hyperparameter values and schedules, which are likely suboptimal, as they were manually tuned on qualitative results obtained from only 20 samples withheld from the test set. A grid search over possible values is currently impractical due to the computational costs of the algorithm, but improvements are definitely possible. This computational cost also represents a general limitation of the algorithm during inference, as the algorithm requires far more GPU cycles than previous methods to generate the artificial libraries at the intermediate stages of the search process. This could be improved with faster image generation models, but will always represent an increase in GPU cycles of several orders of magnitude over direct decoding approaches, as it requires generating many intermediate images in addition to the final image distribution. For details on the computational requirements of our algorithm, see Appendix B.6 and B.7.

\section{Conclusion}

\begin{figure}[h]
\begin{center}
\includegraphics[width=\columnwidth]{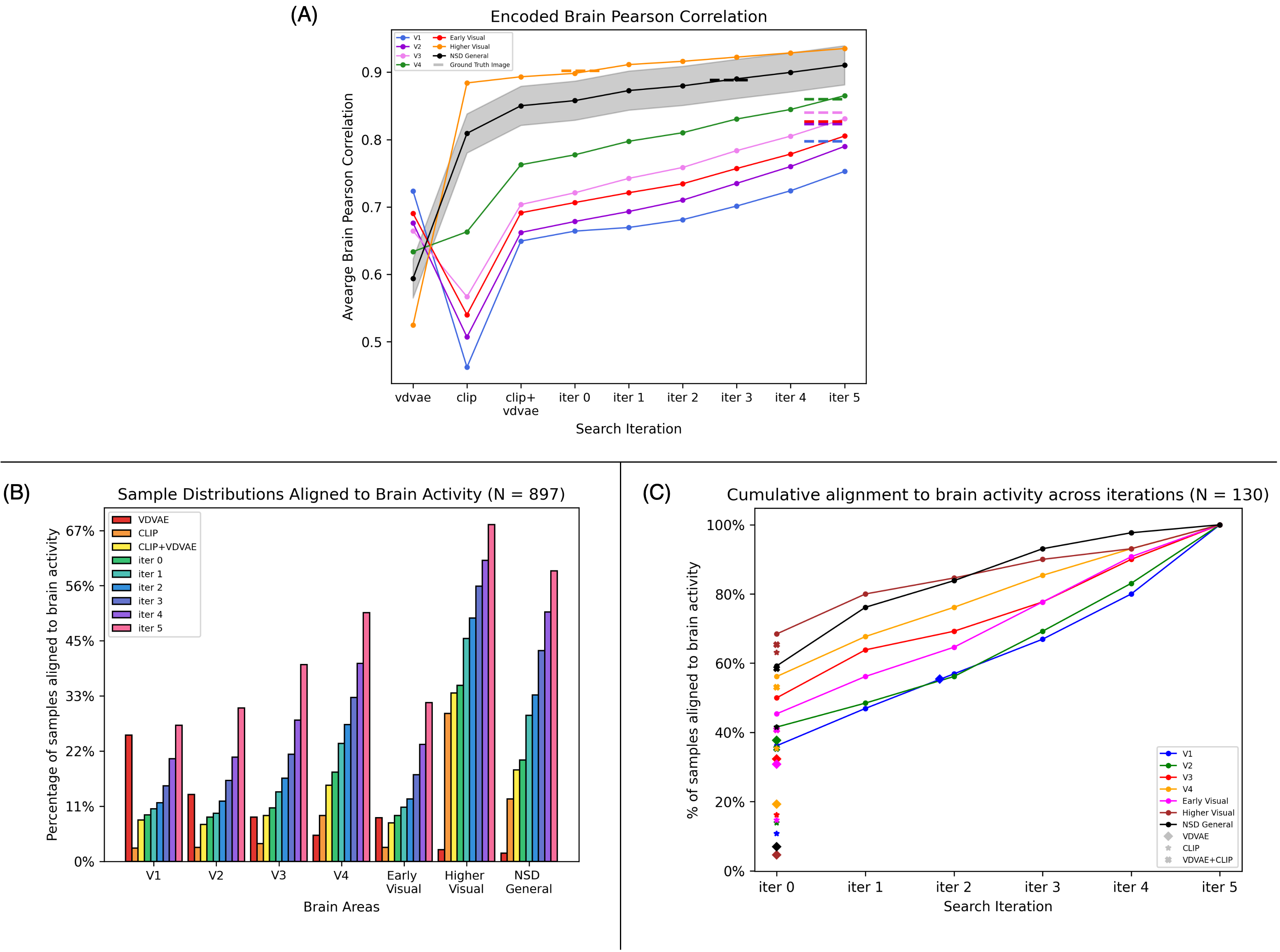}
\end{center}
\caption{(A) Correlation of predicted and actual brain activity (brain correlation) for the top image at each iteration for different ROIs (curves). Dashed lines are the score of the target ground truth image in each ROI. (B) Percentage of samples in the test set for which  reconstructions achieve a brain correlations score at or above parity with the score of the ground truth image (y-axis) for each of the indicated ROIs (x-axis) and across iterations (colors). (C) Cumulative percentage of reconstructions that achieve brain correlations score at or above parity with the score of the ground truth image (y-axis) across iterations (x-axis) and for ``unrefined" reconstructions methods (shapes). This plot uses only samples that reach parity with the score of the ground truth image for all brain areas.} 
\label{figure:plot}
\end{figure}

This work introduces a procedure for reducing the breadth of decoded image distributions while maximizing their alignment to brain activity. We find that this procedure generates high-quality reconstructions that approximate ground truth images in a variety of feature spaces without explicitly minimizing loss in any one of them. A key component of our system is our use of brain-optimized encoding models that yield accurate predictions of brain activity in much of visual cortex \cite{St-Yves_heirarchy}, allowing us to leverage the diversity of representations for natural images across visual cortex. The potential of this new tool is demonstrated in quantifying the invariance in representations across the brain via each brain area’s rate of convergence to parity with the ground truth image, and finding that it sorts brain areas in agreement with complementary measures of invariance explored in other works, such as receptive field size \cite{receptivefields}. We believe image reconstruction algorithms have a pivotal role to play in understanding and interpreting brain activity in visual cortex. By utilizing encoding models to align reconstructions directly with the measured brain activity, this work aims to refine work in this space to a more descriptive and scientific objective that allows these tools to give new insights into the structure and function of various cortical regions.

\section{Acknowledgements}

We would like to thank Stability AI for providing code and pre-trained models for Stable Diffusion. Funding is provided by R01EY023384.

\newpage

\begin{center}
{\huge \textbf{Appendix}}
\end{center}

\appendix
\renewcommand\thefigure{\thesection.\arabic{figure}}    
\renewcommand\thetable{\thesection.\arabic{table}}  
\renewcommand\thefigure{\thesection.\arabic{figure}}    
\renewcommand\thetable{\thesection.\arabic{table}}

\section{Data Withholding}
\setcounter{figure}{0} 
\setcounter{table}{0} 
Previous examination in Tagaki et al. \cite{Takagi2022.11.18.517004} has found that 65 of the images used in the NSD shared1000 test data \cite{Allen2021a}, approximately 7\% of the samples, were present in LAION-5B, which Stable Diffusion and the ViT-H/14 CLIP model utilized for training. Consequently, we excluded these overlapping images from our generated samples, and they were not used in either our quantitative analysis or any of our qualitative examples.

In addition to the samples held out from LAION-5B\cite{schuhmann2022laion5b}, we also withhold the first 20 samples in the test set, which we used to manually tune various hyperparameters, including the number of images generated at each iteration, the number of batches, the number of iterations, the strength and momentum schedules, as well as the $k$ parameter of the Library Assembler. We withhold these samples to avoid overfitting our selected hyperparameters to those data points.

The full list of the 897 samples our quantitative metrics were calculated on will be released along with our code and models in late June 2023.

\section{Architecture details}
\setcounter{figure}{0} 
\setcounter{table}{0} 
\subsection{VDVAE}
For the VDVAE model \cite{vdvae}, we use the pre-trained VDVAE Encoder to extract and concatenate the latent variables for the first 31 layers of the hierarchy, creating a vector with 91,168 dimensions. This allows us to pass images from our large library through the model and collect the VDVAE latents as output, which we can then combine in the Library Assembler. Once the Library Assembler returns a set of averaged VDVAE latents, this set is then reconstructed into an image using the pre-trained VDVAE Decoder. This reconstruction serves as an initial guess for the low-level features of our reconstruction, seeding future stages of the architecture. It's important to note that all VDVAE layers, including both encoder and decoder blocks, are pre-trained and frozen, and we train no additional models for this stage of the pipeline. To convert the image into the $z$ vector format accepted for structural guidance in Stable Diffusion, we pass it through the AutoKL encoder utilized in Stable Diffusion's img2img script. 

\subsection{CLIP}
For the generation of CLIP vectors representing semantic content \cite{radford2021learning}, we utilize an OpenClip implementation of the ViT-H/14 image embedding model \cite{openclip, openclip2}. This generates a CLIPVision vector $c$ with 1024 dimensions. Stable Diffusion 2.1 utilizes this CLIP embedding format, making it a natural choice for this work. It is notably larger than the ViT-L/14 model used in previous versions of Stable Diffusion and other diffusion models, which may contribute to an increase in the semantic fidelity of the reconstructions.

\subsection{Stable Diffusion}
For the diffusion model responsible for image generation, the pre-trained Stable Diffusion unCLIP model from the Diffusers library \cite{stablediffusion} is adopted. This model, released as Stable Diffusion Reimagine, is a modified version of Stable Diffusion 2.1, which has been trained to accept OpenClip ViT-H/14 image embeddings. In this work, the model pipeline is adapted to also receive a latent $z$ vector and a strength parameter $\lambda$ in addition to the $c$ embeddings.

\subsection{Denoising autoencoder}
For the places of our pipeline where we want to compare encoded brain data against measured brain data to generate a brain correlation score, we train a denoising autoencoder to remove some of the additional noise present in the measured brain data. To do this, we encode the original brain data, whose layer size is the voxel count of the nsdgeneral mask, [15724,14278,13039,12682] for subjects 1, 2, 5, and 7, respectively, down to a set of compressed intermediary layers of 5000, 1000, and 500 dimensions. We then decode this layer back up through these layer sizes to the input layer size to produce a denoised representation of the original scan that contains only the relevant signal to our comparison. To train this model, we use the measured brain activity of a sample $\beta$ as input and the encoded $\beta'$ of the ground truth image computed by our hybrid encoding model detailed in section \ref{hybrid} as the target. By training the model in this way, our autoencoder is trained to keep only the signal elements that our encoding model deems relevant, leading to the very high brain correlation scores seen in some metrics, often above 0.9. The fact that the images in this work are tuned to maximize the measured correlation between the outputs of the autoencoder and encoding model may limit the relevance of the brain correlation metric to other works that haven't been optimized for this measure of performance, but it still provides a useful insight into how various reconstructions represent measured brain data.

\subsection{Hybrid encoder}
\label{hybrid}
For the hybrid encoder used in various places in our architecture, we combine the outputs of GNet, a pre-trained interpretable voxel-wise encoding model designed to extract and associate image features with the feature-weighted receptive fields of individual voxels in the visual cortex, and a 3-layer deep neural network that we train to map ViT-H/14 CLIP vectors to fMRI signals. The GNet model was initially trained on the same combined training-validation dataset employed in this research, allowing for the combination of their respective outputs. The CLIP encoding model is sensitive only to semantic features represented in the CLIP embedding, and so largely predicts voxels in areas outside the early visual cortex, while GNet provides predictions that are very accurate within the early visual cortex but less sensitive to higher regions corresponding to semantic representations of visual content. This discrepancy serves as the motivation for combining their outputs to get a more comprehensive estimate of brain activity. Different variations of combining the outputs of each method were briefly tested, including weighting individual voxel predictions by the relative prediction accuracies of each model; however, this did not lead to an increase in prediction accuracy, so a simple averaging procedure was used to combine outputs. More rigorous experiments for best combining the outputs of different encoding models could be done in the future to expand upon this implementation, such as the work done in \cite{Lin2023.04.23.537988}. 

\subsection{Library Assembler}
The library used for this work is a subset of the 73,000 COCO \cite{microsoftcoco} images used as stimuli in the Natural Scenes Dataset \cite{Allen2021a}. Each subject of the experiment viewed up to 10,000 of these images, and so when computing the initial latent variables for a subject, the 63,000 images not seen by that subject at any point during the experiment are used. This means that the 63,000 images used in the Library Assembler stage will be different for each subject, but should contain a very similar distribution of images and semantic classes.

For the $k$ parameter used in the Library Assembler, we performed a grid search over possible values of $k$ on the 20 samples held out from the test set to find values that sufficiently maximized quantitative and qualitative comparisons against the latent variables extracted from the corresponding ground truth images. In this analysis, we determined that a value of $\approx{25}$ works well for the VDVAE latent configuration, while a value of $\approx{100}$ is optimal for the CLIP configuration. Notably, CLIP vectors lie in a Euclidean vector space, meaning you can average as many of them together as you like without significant loss in reconstruction quality. The optimal $k$ value in this instance should correspond to the number of images in our large library (n=63,000) that fall within close proximity of the semantic class of the seen image. The VDVAE latents, however, do not fall within a neat Euclidean vector space. While they can still be averaged together, the larger the set of image latents averaged together, the greater the distortion of the produced image. The $k$ value of 25, in this case, was selected to strike a qualitative balance between not overfitting our initial structural guess to a single library image, and not overly distorting our initial structural guess by averaging too many latents together.

The only GPU operations required for the Library Assembler are the inference stage of the encoding model used to create $\beta'$ values for each image, and the inference stage of the denoising autoencoder to create $\hat{\beta}$s for each of your measured samples. This makes the GPU overhead of the architecture quite reasonable, taking about 1 hour for both of these steps for our n=63,000 library on an NVIDIA A100 with ~17GB of VRAM being utilized. The other major computation requirement for this architecture is its memory requirement. The current implementation loads all 63,000 latent variables corresponding to library images into memory to provide fast access to them during the sorting stage. At 0.7mb for our largest latent variable, the VDVAE latents, that amounts to 46GB of memory usage. The inference stage of the Library Assembler, where it performs the search and averaging steps, takes about 10 minutes per subject to compute latent variables for the 897 test samples.

\subsection{Stochastic Search}
Within the SCS algorithm, the strength parameter $\lambda$ at each iteration is calculated using the equation $\lambda = 0.92-0.3(\frac{i+1}{6})^3$, where $i$ denotes the current iteration number. The momentum parameter $\gamma$ is calculated by the equation $\gamma = 0.2(\frac{i+1}{6})^2$. These schedules, as well as the numerous hyperparameters of the algorithm, including $i$, the number of iterations, $n$ the number of images generated at each iteration, and $n_b$ the number of images in each batch, were all tuned to their respective values through qualitative testing on the first 20 held out test samples described in Section A.

The Stochastic Search algorithm presents a very large GPU computing requirement. At each iteration of the search, we use a diffusion model to generate hundreds of artificial library images and an encoding model to encode each of them into $\beta'$s to be scored against our denoised $\hat{\beta}$s. By far, the largest cost comes in the diffusion process of image generation. Each image takes about 3 seconds to generate on an NVIDIA A100. At 100 images per iteration and 6 iterations, that is 30 minutes of computing time for just the image generations of a single test sample. To help address this, we implement an optimization to reduce the number of samples generated in each batch as the strength parameter decreases, as we need fewer samples to represent the same variance in our distribution. This optimization is calculated by taking the base number of branch samples ($25$) and multiplying it by our current strength parameter value $\lambda$, so the number of batch samples is calculated by $n_b = 25\lambda$. With these optimizations, we reduce the number of images generated by 37\% and reduce the time requirement for the diffusion step to about 20 minutes per sample. The additional overhead of the encoding model predictions for those images as well as other file operation steps, brings our total compute time for a single sample to 31.5 minutes on an NVIDIA A100. Our total test set generations were computed across 4 A100s, one per subject, for a total of 19 days.

\section{Quantitative Metrics}
\setcounter{figure}{0} 
\setcounter{table}{0} 

\subsection{Details for Table 1}
 The metrics for Ozcelik et al. \cite{ozcelik2023braindiffuser} were calculated using 5 repetitions of the results generated using their open-source implementation. Ozcelik et al. also provided a single repetition of their results, which we verified had very similar metrics to our 5-fold reproduction. The results for Tagaki et al. \cite{Takagi2022.11.18.517004} were calculated on 5 repetitions of their samples provided by the authors. The authors of Lu et al. \cite{lu2023minddiffuser} and Gu et al. \cite{lu2023minddiffuser} only provided one repetition of their reconstructions, and do not have open-source code, and so their metrics are calculated only across the single image for each sample that the authors provided us. This difference should not have a large effect on their metrics, besides making their results slightly less confident.

The CLIP model used for the quantitative values in both the CLIP(2-way) and CLIP(cos) metrics was the official ViT-L/14 model (768 dimensions) released by OpenAI. This is notably different than the OpenCLIP ViT-H/14 model (1024 dimensions) used for our pipeline but is a more commonly used model for calculating CLIP metrics in other papers such as Ozcelik et al\cite{ozcelik2023braindiffuser}. and Tagaki et al.\cite{Takagi2022.11.18.517004}, and so was used for the sake of consistency with other works. It is different than the CLIP model used in Lu et al. \cite{lu2023minddiffuser}, which appears to use a ViT-B/16 model (512 dimensions). Out of curiosity, we tested these other CLIP models to see if they would produce a difference in quantitative benchmarks and confirmed that the results vary drastically in scale between models, confirming the need for consistency in the calculation of this metric.

Another notable item is that for all data reported in this paper, the SSIM metric was calculated using version 0.20.0 of the Scikit-Image python package, which is the newest version at the time of this work. This version carries a significant change in functionality from previous versions of the package used to calculate the metric in previous works such as Ozcelik et al., Lu et al., and Gu et al. \cite{ozcelik2023braindiffuser, lu2023minddiffuser, gu2023decoding}, so the SSIM metrics reported here may not align with the numbers reported in their papers.

\subsection{Multi-subject results}
The authors of Lu et al. \cite{lu2023minddiffuser} did not produce results on additional subjects beyond subject 1, and so their results are excluded from Table C.1. The remaining results are averaged across subjects 1, 2, 5, and 7, in accordance with the sample repetition methodology detailed in Section C.1.

\begin{table}[!htb]
\centering
\resizebox{\columnwidth}{!}{%
\begin{tabular}{lccccc}
\hline
\multicolumn{1}{|c|}{Methods}                   & \multicolumn{5}{c|}{Low Level Structural Metrics}                                                                                                                              \\ \cline{2-6} 
\multicolumn{1}{|c|}{}                          & \multicolumn{1}{c|}{PixCorr↑}     & \multicolumn{1}{c|}{SSIM↑}      & \multicolumn{1}{c|}{AlexNet(2)↑}   & \multicolumn{1}{c|}{AlexNet(5)↑} & \multicolumn{1}{c|}{AlexNet(7)↑} \\ \hline
\multicolumn{1}{|l|}{Second Sight (ours)}  & \multicolumn{1}{c|}{.156}
& \multicolumn{1}{c|}{\textbf{.285}}  & \multicolumn{1}{c|}{88.4\%}          & \multicolumn{1}{c|}{93.5\%}     & \multicolumn{1}{c|}{93.1\%}           \\
\multicolumn{1}{|l|}{Best Library Image (ours)} & \multicolumn{1}{c|}{.124} & \multicolumn{1}{c|}{.252} & \multicolumn{1}{c|}{84.2\%} & \multicolumn{1}{c|}{91.6\%}    & \multicolumn{1}{c|}{91.9\%}           \\

\multicolumn{1}{|l|}{Ozcelik et al.}  & \multicolumn{1}{c|}{\textbf{.253}}   & \multicolumn{1}{c|}{.284} & \multicolumn{1}{c|}{\textbf{94.1\%}}    & \multicolumn{1}{c|}{\textbf{96.3\%}}  & \multicolumn{1}{c|}{\textbf{95.6\%}}  \\

\multicolumn{1}{|l|}{Tagaki et al.}  & \multicolumn{1}{c|}{.213}            & \multicolumn{1}{c|}{.213}      & \multicolumn{1}{c|}{71.0\%}          & \multicolumn{1}{c|}{74.0\%}    & \multicolumn{1}{c|}{73.9\%}           \\

\multicolumn{1}{|l|}{Gu et al.} & \multicolumn{1}{c|}{.141} & \multicolumn{1}{c|}{.276} & \multicolumn{1}{c|}{78.2\%} & \multicolumn{1}{c|}{89.0\%}   & \multicolumn{1}{c|}{92.5\%}   \\ \hline
                                                & \multicolumn{1}{l}{}              & \multicolumn{1}{l}{}            & \multicolumn{1}{l}{}               & \multicolumn{1}{l}{}             & \multicolumn{1}{l}{}             \\ \hline
\multicolumn{1}{|c|}{Methods}                   & \multicolumn{5}{c|}{High Level Semantics Metrics}                                                                                                                              \\ \cline{2-6} 
\multicolumn{1}{|c|}{}                          & \multicolumn{1}{c|}{CLIP(2-way)↑} & \multicolumn{1}{c|}{CLIP(cos)↑} & \multicolumn{1}{c|}{Inception V3↑} & \multicolumn{1}{c|}{EffNet-B↓}   & \multicolumn{1}{c|}{SwAV↓}       \\ \hline
\multicolumn{1}{|l|}{Second Sight (ours)}  & \multicolumn{1}{c|}{87.0\%}   & \multicolumn{1}{c|}{\textbf{.655}} & \multicolumn{1}{c|}{82.0\%}           & \multicolumn{1}{c|}{.792}           & \multicolumn{1}{c|}{.435}           \\

\multicolumn{1}{|l|}{Best Library Image (ours)} & \multicolumn{1}{c|}{84.4\%}       & \multicolumn{1}{c|}{.645}          & \multicolumn{1}{c|}{80.5\%}             & \multicolumn{1}{c|}{.818}           & \multicolumn{1}{c|}{.458}           \\

\multicolumn{1}{|l|}{Ozcelik et al.}            & \multicolumn{1}{c|}{\textbf{91.7\%}}   & \multicolumn{1}{c|}{.653}          & \multicolumn{1}{c|}{\textbf{87.5\%}}    & \multicolumn{1}{c|}{\textbf{.774}}  & \multicolumn{1}{c|}{\textbf{.423}}  \\

\multicolumn{1}{|l|}{Tagaki et al.} & \multicolumn{1}{c|}{64.1\%}            & \multicolumn{1}{c|}{.564}    & \multicolumn{1}{c|}{63.2\%}             & \multicolumn{1}{c|}{.957}   & \multicolumn{1}{c|}{.671}           \\

\multicolumn{1}{|l|}{Gu et al.} & \multicolumn{1}{c|}{77.7\%}   & \multicolumn{1}{c|}{.587}      & \multicolumn{1}{c|}{78.6\%}             & \multicolumn{1}{c|}{.870}           & \multicolumn{1}{c|}{.459}           \\ \hline
\end{tabular}%
}
\caption{Quantitative comparison against past reconstruction methods on subjects 1, 2, 5, and 7. For each measure, the best value is in bold. For EffNet-B and SwAV distances, lower is better. Higher is better for all other metrics. This is indicated by the arrow pointing up or down.}
\end{table}

\begin{table}[!htb]
\centering
\resizebox{\textwidth}{!}{%
\begin{tabular}{|l|cccccc|}
\hline
\multicolumn{1}{|c|}{Methods} & \multicolumn{6}{c|}{Brain Correlation Metrics}                                                                                                        \\ \cline{2-7} 
\multicolumn{1}{|c|}{}        & \multicolumn{1}{c|}{V1} & \multicolumn{1}{c|}{V2} & \multicolumn{1}{c|}{V3} & \multicolumn{1}{c|}{V4} & \multicolumn{1}{c|}{Higher Vis} & nsdgeneral \\ \hline

Second Sight (ours)   & \multicolumn{1}{c|}{\textbf{.795}} 
& \multicolumn{1}{c|}{\textbf{.814}}  & \multicolumn{1}{c|}{\textbf{.845}}
& \multicolumn{1}{c|}{\textbf{.866}}  & \multicolumn{1}{c|}{\textbf{.937}}   & \textbf{.92}           \\

Best Library Image (ours)     & \multicolumn{1}{c|}{.573}  & \multicolumn{1}{c|}{.588}  & \multicolumn{1}{c|}{.606}  & \multicolumn{1}{c|}{.612}  & \multicolumn{1}{c|}{.656}          & .65           \\

Ozcelik et al.                & \multicolumn{1}{c|}{.716}  & \multicolumn{1}{c|}{.715}  & \multicolumn{1}{c|}{.739}  & \multicolumn{1}{c|}{.766}  & \multicolumn{1}{c|}{.866}          & .844           \\

Tagaki et al. & \multicolumn{1}{c|}{.311}  & \multicolumn{1}{c|}{.260}  
              & \multicolumn{1}{c|}{.264}  & \multicolumn{1}{c|}{.306}  & \multicolumn{1}{c|}{.360}  & .353           \\

Gu et al.  & \multicolumn{1}{c|}{.482}  & \multicolumn{1}{c|}{.499}  & \multicolumn{1}{c|}{.524}  & \multicolumn{1}{c|}{.574}  & \multicolumn{1}{c|}{.729}          & .691          \\ \hline
\end{tabular}%
}
\caption{Evaluation of brain correlation scores against past reconstruction methods on subjects 1, 2, 5 and 7. Evaluating how well the reconstructed images represent activity in various brain regions according to our brain correlation metric.}
\label{table:braincorr}
\end{table}

\section{Ablation studies}
\setcounter{figure}{0} 
\setcounter{table}{0} 

Here we present an ablation study of how the various parts of our Second Sight architecture perform in isolation with regard to subject 1. In the qualitative analysis found in Figures \ref{figure:itercomp} and \ref{figure:itercomp2}, the Only VDVAE model demonstrated potential for reproducing low-level details, such as shapes and layouts, of the original ground truth images. However, these reconstructions often resembled hazy outlines, falling short of recreating realistic images. On the contrary, the Only CLIP model performed well in high-level properties, which are essentially the semantics of the ground truth images. However, it struggled with capturing the accurate placement and layout of objects within these images. The combination of both CLIP and VDVAE outperforms both standalone models by capturing elements of both high-level and low-level information. This combined approach, thus, serves as our initial benchmark for the reconstruction process. As the iterative search process begins, the accuracy of the reconstructions improves, aligning more closely with the reference images, thereby enhancing the accuracy of the reconstructions. 

In our quantitative analysis found in Tables \ref{table:ablation} and \ref{table:ablation2}, we observe similar conclusions. The first variant, using only the low-level structural guess provided by the VDVAE latents, excelled in the low-level metrics, specifically PixCorr, SSIM, and AlexNet(2). However, it faltered in high-level measures and in terms of brain correlation outside of V1. The second variant, Only CLIP, yielded underwhelming results across all measures, including low-level, high-level, and brain correlation metrics. The hybrid approach of CLIP+VDVAE demonstrated intermediate results, approximating the average performance of the two isolated methods. These outcomes provide a benchmark for image reconstruction in the future iterative stages, which aim to improve them. Observing these iterations, PixCorr and SSIM performance deteriorates as the search progresses, while AlexNet(2) maintains a stable performance. Most notably, the performances of AlexNet(5) and AlexNet(7), as well as the high-level metrics of CLIP(2-way), Inception V3, and SwAV ,note an increase in performance as the search advances. However, CLIP cosine and EffNet-B performance initially improve before declining in later stages. At the fifth iteration, a significant drop in performance is noticeable across multiple metrics, including AlexNet(2), AlexNet(5), AlexNet(7), CLIP(2-way), Inception V3, EffNet-B, and SwAV. This decline is not observed in the brain correlation, as observed in the brain correlation measurements, reaffirming the conclusion that the brain's representation of seen images is abstracted away from the original seen image, and can be adequately reproduced by many different images. Overall, the image reconstruction quality initially improves as the search progresses, reaches an optimal point, and then declines as the image begins to align more closely with the subject's brain responses. This study provides insights into the nuanced effects of the search iterations on image reconstruction quality, as well as how the different parts of our pipeline contribute to our final results.

\begin{figure}[!htb]
\begin{center}
\includegraphics[width=\columnwidth]{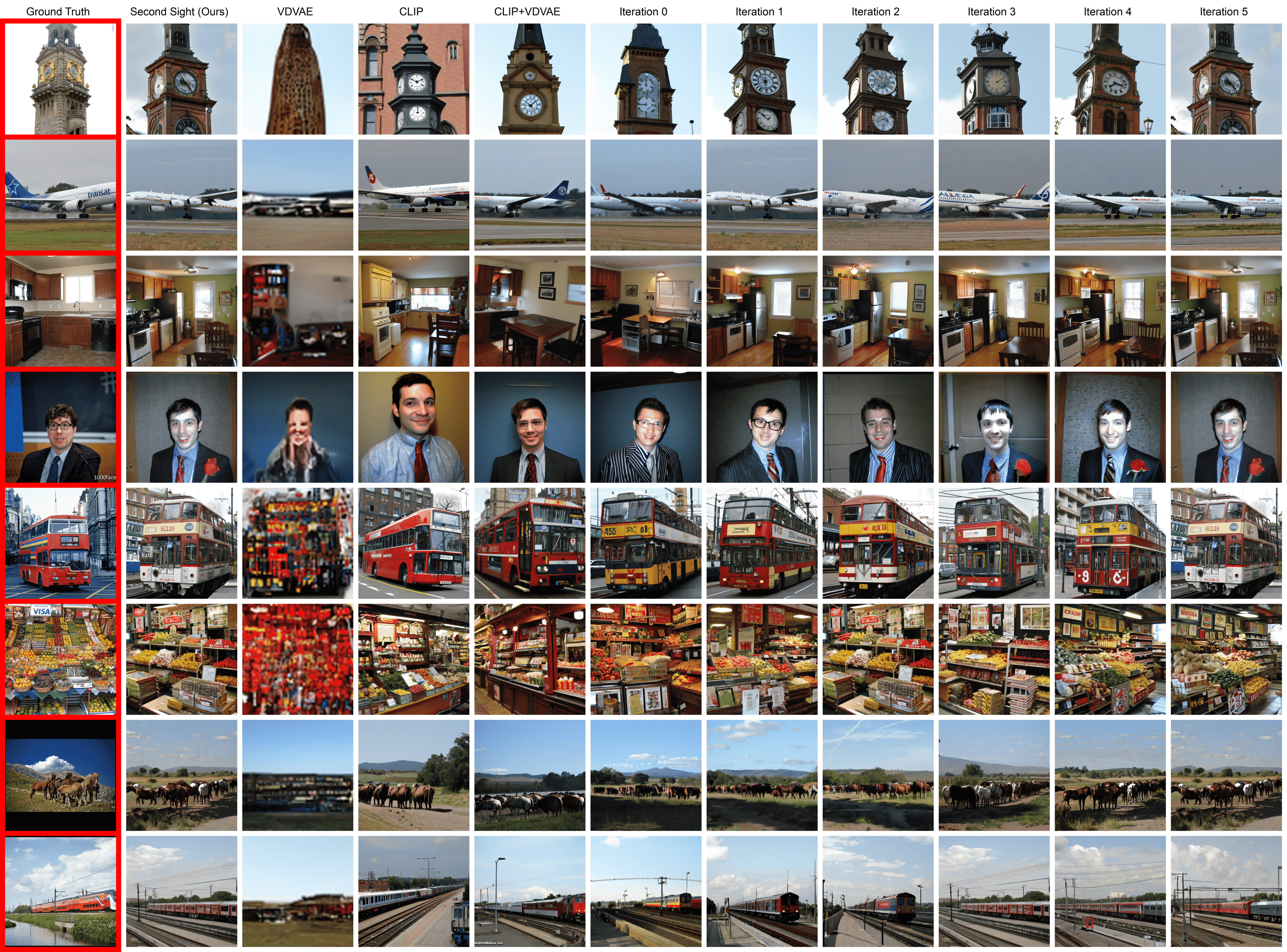}
\end{center}
\caption{Qualitative ablation study of our Second Sight architecture on subject 1. The samples are the same as in Figure 2 to allow for easy comparison. The first column is the ground truth image, indicated by the red background, while the remaining columns represent the output at various stages of the Second Sight algorithm.} 
\label{figure:itercomp}
\end{figure}

\begin{figure}[!htb]
\begin{center}
\includegraphics[width=\columnwidth]{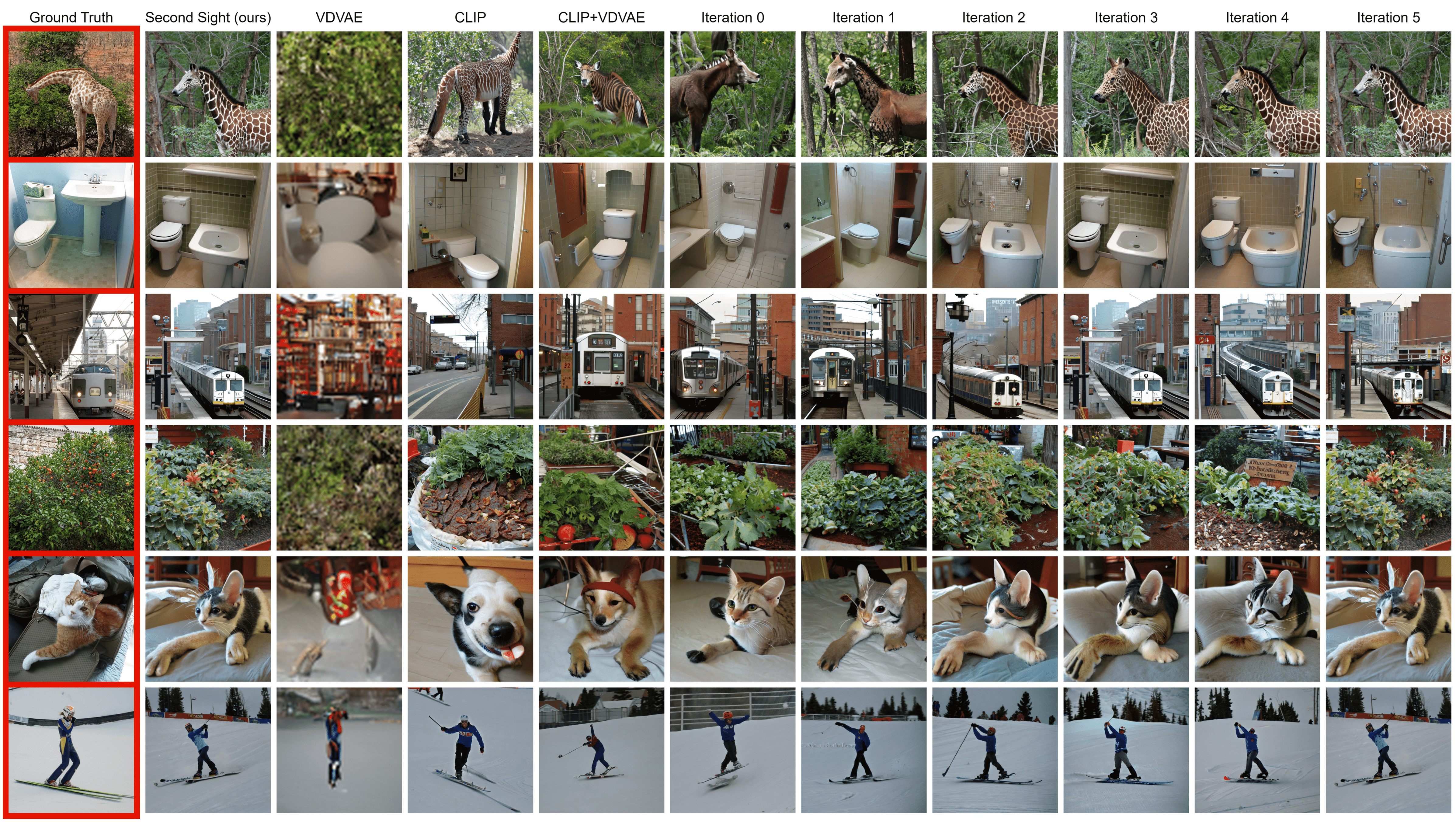}
\end{center}
\caption{Some additional ablation study samples with comparison across stages of our pipeline, ground truth images are indicated by a red background. These samples were picked as examples that specifically demonstrate the effects of the iterative refinement capabilities of our architecture.} 
\label{figure:itercomp2}
\end{figure}

\begin{table}[!htb]
\centering
\resizebox{\columnwidth}{!}{%
\begin{tabular}{lccccc}
\hline
\multicolumn{1}{|c|}{Methods}             & \multicolumn{5}{c|}{Low Level Structural Metrics}                                                                                                                              \\ \cline{2-6} 
\multicolumn{1}{|c|}{}                    & \multicolumn{1}{c|}{PixCorr↑}     & \multicolumn{1}{c|}{SSIM↑}      & \multicolumn{1}{c|}{AlexNet(2)↑}   & \multicolumn{1}{c|}{AlexNet(5)↑} & \multicolumn{1}{c|}{AlexNet(7)↑} \\ \hline

\multicolumn{1}{|l|}{Only VDVAE} & \multicolumn{1}{c|}{\textbf{.252}}            & \multicolumn{1}{c|}{\textbf{.350}}    & \multicolumn{1}{c|}{\textbf{92.4\%}}         & \multicolumn{1}{c|}{91.7\%}  & \multicolumn{1}{c|}{87.0\%}       \\

\multicolumn{1}{|l|}{Only CLIP} & \multicolumn{1}{c|}{.083}            & \multicolumn{1}{c|}{.268}   & \multicolumn{1}{c|}{80.6\%}           
 & \multicolumn{1}{c|}{89.8\%} & \multicolumn{1}{c|}{92.5\%}      \\

\multicolumn{1}{|l|}{CLIP+VDVAE}   & \multicolumn{1}{c|}{.220}  & \multicolumn{1}{c|}{.310} & \multicolumn{1}{c|}{90.6\%}   & \multicolumn{1}{c|}{93.7\%}  & \multicolumn{1}{c|}{93.4\%}  \\

\multicolumn{1}{|l|}{Iteration 0} & \multicolumn{1}{c|}{.219}            & \multicolumn{1}{c|}{.310}     & \multicolumn{1}{c|}{91.1\%}          & \multicolumn{1}{c|}{93.8\%}   & \multicolumn{1}{c|}{93.5\%}           \\

\multicolumn{1}{|l|}{Iteration 1} & \multicolumn{1}{c|}{.206}           & \multicolumn{1}{c|}{.296}     & \multicolumn{1}{c|}{90.8\%}         
 & \multicolumn{1}{c|}{94.3\%}   & \multicolumn{1}{c|}{93.9\%}   \\

\multicolumn{1}{|l|}{Iteration 2} & \multicolumn{1}{c|}{.197}           & \multicolumn{1}{c|}{.293}    & \multicolumn{1}{c|}{90.6\%}           & \multicolumn{1}{c|}{94.5\%}  & \multicolumn{1}{c|}{93.9\%}     \\

\multicolumn{1}{|l|}{Iteration 3} & \multicolumn{1}{c|}{.186}          
& \multicolumn{1}{c|}{.288}    & \multicolumn{1}{c|}{90.7\%}           & \multicolumn{1}{c|}{94.5\%}  & \multicolumn{1}{c|}{94.0\%}       \\

\multicolumn{1}{|l|}{Iteration 4} & \multicolumn{1}{c|}{.181}          & \multicolumn{1}{c|}{.285}    & \multicolumn{1}{c|}{91.8\%}           & \multicolumn{1}{c|}{94.7\%}  & \multicolumn{1}{c|}{94.1\%}       \\

\multicolumn{1}{|l|}{Iteration 5} & \multicolumn{1}{c|}{.136}          & \multicolumn{1}{c|}{.281}    & \multicolumn{1}{c|}{76.9\%}           
& \multicolumn{1}{c|}{87.1\%}  & \multicolumn{1}{c|}{90.4\%}   \\

\multicolumn{1}{|l|}{Second Sight (ours)}  & \multicolumn{1}{c|}{.185}          & \multicolumn{1}{c|}{.288}   & \multicolumn{1}{c|}{91.7\%}          & \multicolumn{1}{c|}{\textbf{95.0\%}}   & \multicolumn{1}{c|}{\textbf{94.2\%}}         \\ \hline
                                          & \multicolumn{1}{l}{}              & \multicolumn{1}{l}{}            & \multicolumn{1}{l}{}               & \multicolumn{1}{l}{}             & \multicolumn{1}{l}{}             \\ \hline
\multicolumn{1}{|c|}{Methods}  & \multicolumn{5}{c|}{High Level Semantics Metrics}                                                                                                                              \\ \cline{2-6} 
\multicolumn{1}{|c|}{}                    & \multicolumn{1}{c|}{CLIP(2-way)↑} & \multicolumn{1}{c|}{CLIP(cos)↑} & \multicolumn{1}{c|}{Inception V3↑} & \multicolumn{1}{c|}{EffNet-B↓}   & \multicolumn{1}{c|}{SwAV↓}       \\ \hline

\multicolumn{1}{|l|}{Only VDVAE} & \multicolumn{1}{c|}{67.2\%}            & \multicolumn{1}{c|}{.571}   & \multicolumn{1}{c|}{70.1\%}             & \multicolumn{1}{c|}{.933}  & \multicolumn{1}{c|}{.600}           \\

\multicolumn{1}{|l|}{Only CLIP}  & \multicolumn{1}{c|}{87.8\%}            & \multicolumn{1}{c|}{.656} & \multicolumn{1}{c|}{82.2\%}             & \multicolumn{1}{c|}{.792}  & \multicolumn{1}{c|}{.432}           \\

\multicolumn{1}{|l|}{CLIP+VDVAE} & \multicolumn{1}{c|}{87.7\%}   & \multicolumn{1}{c|}{.661} & \multicolumn{1}{c|}{83.0\%}    & \multicolumn{1}{c|}{.788}  & \multicolumn{1}{c|}{.425}  \\

\multicolumn{1}{|l|}{Iteration 0}& \multicolumn{1}{c|}{87.7\%}         & \multicolumn{1}{c|}{.665}    & \multicolumn{1}{c|}{82.8\%}             & \multicolumn{1}{c|}{.786} & \multicolumn{1}{c|}{.423}         
\\

\multicolumn{1}{|l|}{Iteration 1} & \multicolumn{1}{c|}{88.1\%}        & \multicolumn{1}{c|}{.665}    & \multicolumn{1}{c|}{83.1\%}             & \multicolumn{1}{c|}{.783} & \multicolumn{1}{c|}{.421}           \\

\multicolumn{1}{|l|}{Iteration 2} & \multicolumn{1}{c|}{88.1\%}        
 & \multicolumn{1}{c|}{\textbf{.666}}   & \multicolumn{1}{c|}{83.1\%}             & \multicolumn{1}{c|}{.780} & \multicolumn{1}{c|}{.420}           \\

\multicolumn{1}{|l|}{Iteration 3} & \multicolumn{1}{c|}{88.2\%}         & \multicolumn{1}{c|}{.663}     & \multicolumn{1}{c|}{83.3\%}             & \multicolumn{1}{c|}{.779}  & \multicolumn{1}{c|}{.421}           \\

\multicolumn{1}{|l|}{Iteration 4} & \multicolumn{1}{c|}{88.0\%}         & \multicolumn{1}{c|}{.662}     & \multicolumn{1}{c|}{83.8\%}             & \multicolumn{1}{c|}{.780}  & \multicolumn{1}{c|}{.421}           \\

\multicolumn{1}{|l|}{Iteration 5} & \multicolumn{1}{c|}{75.3\%}        & \multicolumn{1}{c|}{.660}     & \multicolumn{1}{c|}{75.8\%}            & \multicolumn{1}{c|}{.893}    & \multicolumn{1}{c|}{.479}           \\

\multicolumn{1}{|l|}{Second Sight (ours)}& \multicolumn{1}{c|}{\textbf{88.3\%}}    & \multicolumn{1}{c|}{.664} & \multicolumn{1}{c|}{\textbf{84.7\%}}     & \multicolumn{1}{c|}{\textbf{.776}}          & \multicolumn{1}{c|}{\textbf{.418}}     \\ \hline
\end{tabular}%
}
\caption{Quantitative ablation study on subject 1. For each measure, the best value is in bold. For EffNet-B and SwAV distances, lower is better. Higher is better for all other metrics. This is indicated by the arrow pointing up or down.}
\label{table:ablation}
\end{table}

\begin{table}[!htb]
\centering
\resizebox{\textwidth}{!}{%
\begin{tabular}{|l|cccccc|}
\hline
\multicolumn{1}{|c|}{Methods} & \multicolumn{6}{c|}{Brain Correlation Metrics}                                                                                                        \\ \cline{2-7} 
\multicolumn{1}{|c|}{}        & \multicolumn{1}{c|}{V1} & \multicolumn{1}{c|}{V2} & \multicolumn{1}{c|}{V3} & \multicolumn{1}{c|}{V4} & \multicolumn{1}{c|}{Higher Vis} & nsdgeneral \\ \hline

Only VDVAE   & \multicolumn{1}{c|}{.724}  & \multicolumn{1}{c|}{.676}  
            & \multicolumn{1}{c|}{.665}  & \multicolumn{1}{c|}{.634}  
            & \multicolumn{1}{c|}{.525}  & .594           \\

Only CLIP  & \multicolumn{1}{c|}{.462}  & \multicolumn{1}{c|}{.507}    
            & \multicolumn{1}{c|}{.567}  & \multicolumn{1}{c|}{.663}  
            & \multicolumn{1}{c|}{.884}  & .809           \\
            
CLIP+VDVAE  & \multicolumn{1}{c|}{.649}  & \multicolumn{1}{c|}{.662}  
            & \multicolumn{1}{c|}{.704}  & \multicolumn{1}{c|}{.763}  
            & \multicolumn{1}{c|}{.893}  & .850           \\

Iteration 0 & \multicolumn{1}{c|}{.664}  & \multicolumn{1}{c|}{.678}  
            & \multicolumn{1}{c|}{.721}  & \multicolumn{1}{c|}{.777}  
            & \multicolumn{1}{c|}{.898}  & .858           \\

Iteration 1 & \multicolumn{1}{c|}{.670}  & \multicolumn{1}{c|}{.693}  
            & \multicolumn{1}{c|}{.742}  & \multicolumn{1}{c|}{.798}  
            & \multicolumn{1}{c|}{.911}  & .873           \\

Iteration 2 & \multicolumn{1}{c|}{.681}  & \multicolumn{1}{c|}{.710}  
            & \multicolumn{1}{c|}{.759}  & \multicolumn{1}{c|}{.810}  
            & \multicolumn{1}{c|}{.916}  & .880           \\

Iteration 3 & \multicolumn{1}{c|}{.701}  & \multicolumn{1}{c|}{.735}  
            & \multicolumn{1}{c|}{.784}  & \multicolumn{1}{c|}{.831}  
            & \multicolumn{1}{c|}{.922}  & .890           \\

Iteration 4 & \multicolumn{1}{c|}{.724}  & \multicolumn{1}{c|}{.760}  
            & \multicolumn{1}{c|}{.805}  & \multicolumn{1}{c|}{.845}  
            & \multicolumn{1}{c|}{.929}  & .900           \\

Iteration 5 & \multicolumn{1}{c|}{\textbf{.753}}  & \multicolumn{1}{c|}{\textbf{.790}} & \multicolumn{1}{c|}{\textbf{.831}} & \multicolumn{1}{c|}{\textbf{.865}}  & \multicolumn{1}{c|}{\textbf{.935}} & \textbf{.910}  \\

Second Sight (ours) & \multicolumn{1}{c|}{.739} & \multicolumn{1}{c|}{.774}  & \multicolumn{1}{c|}{.817}  & \multicolumn{1}{c|}{.854}  
        & \multicolumn{1}{c|}{.932}  & .905   \\ \hline

\end{tabular}%
}
\caption{Ablation study of brain correlation scores on subject 1. Evaluating how well the reconstructed images represent activity in various brain regions according to our hybrid encoding model.}
\label{table:ablation2}
\end{table}

\section{Additional samples}
\setcounter{figure}{0} 
\setcounter{table}{0} 

\begin{figure}[!htb]
    \centering
    \includegraphics[width=0.48\textwidth]{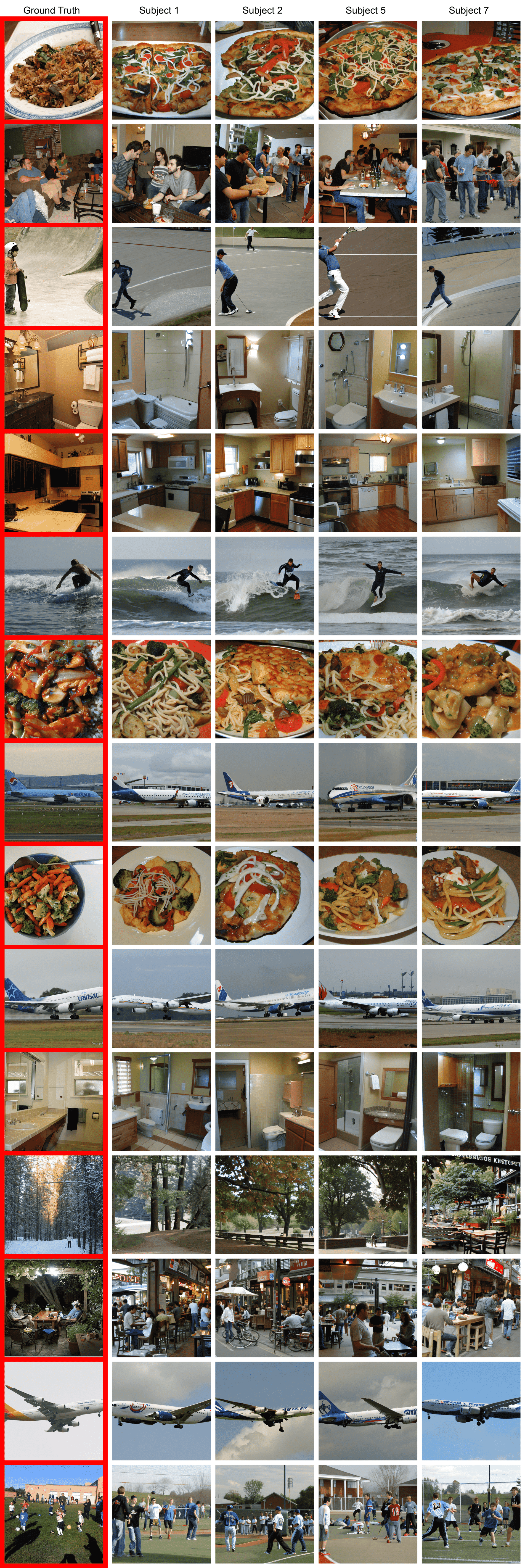}
    \includegraphics[width=0.48\textwidth]{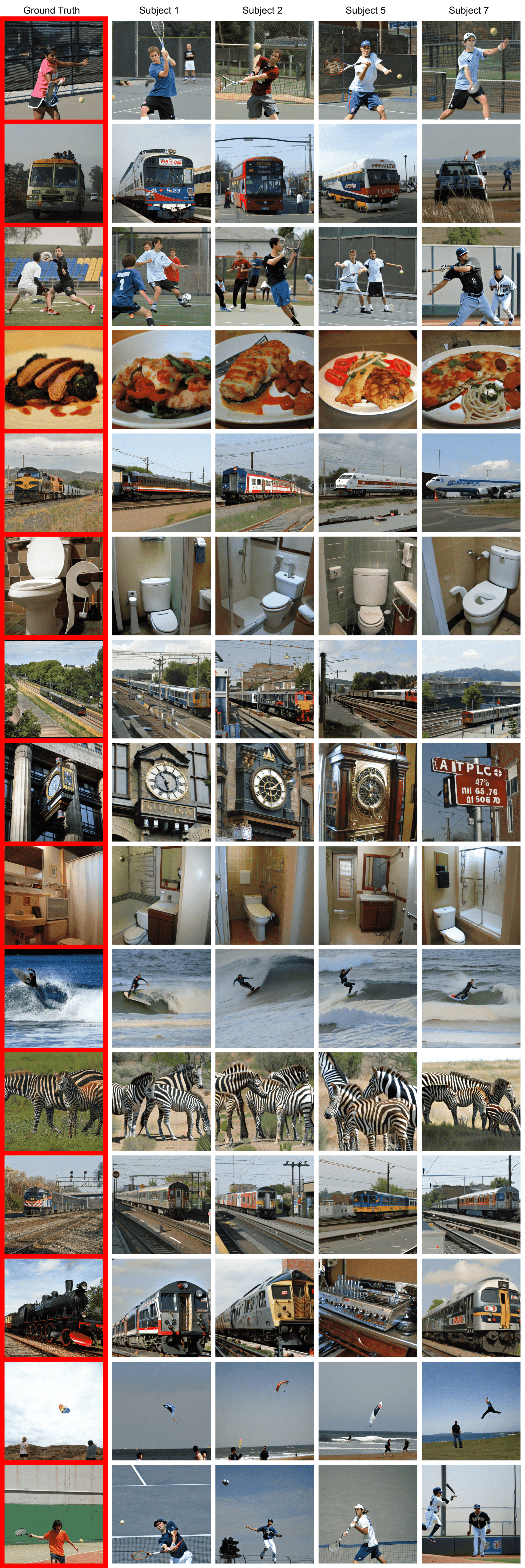}
    \caption{Some additional samples with comparison across subjects, ground truth images are indicated by a red background.}
    \label{fig:appendix1}
\end{figure}

\begin{figure}[!htb]
    \centering
    \includegraphics[width=0.48\textwidth]{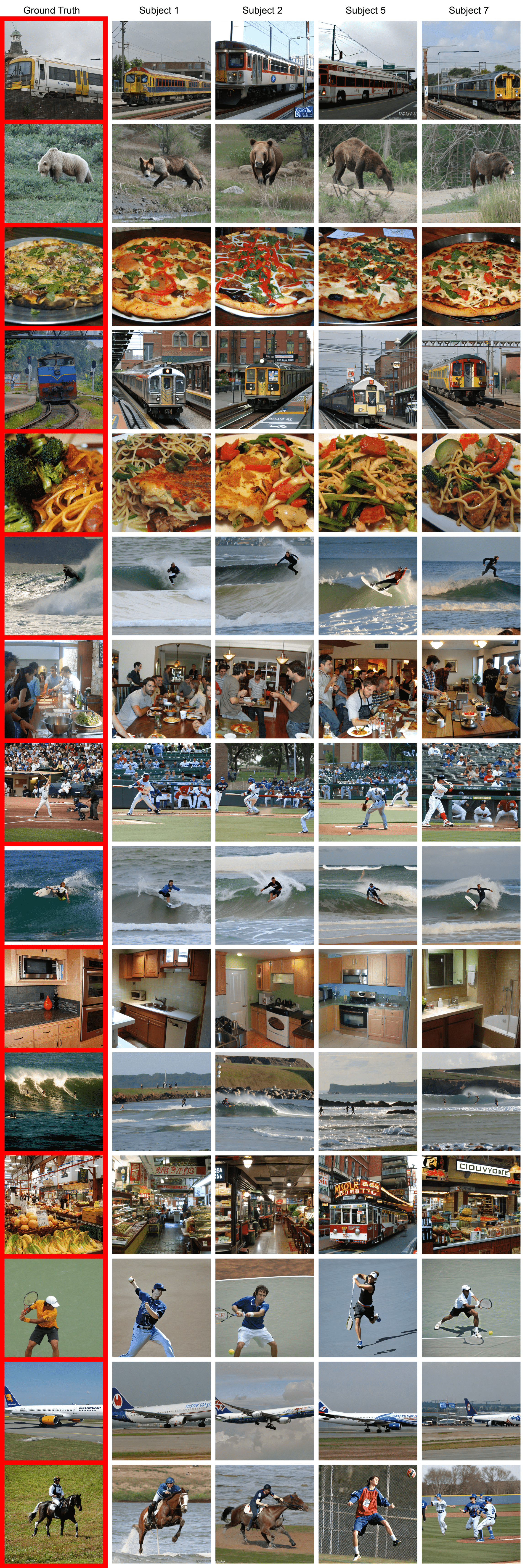}
    \includegraphics[width=0.48\textwidth]{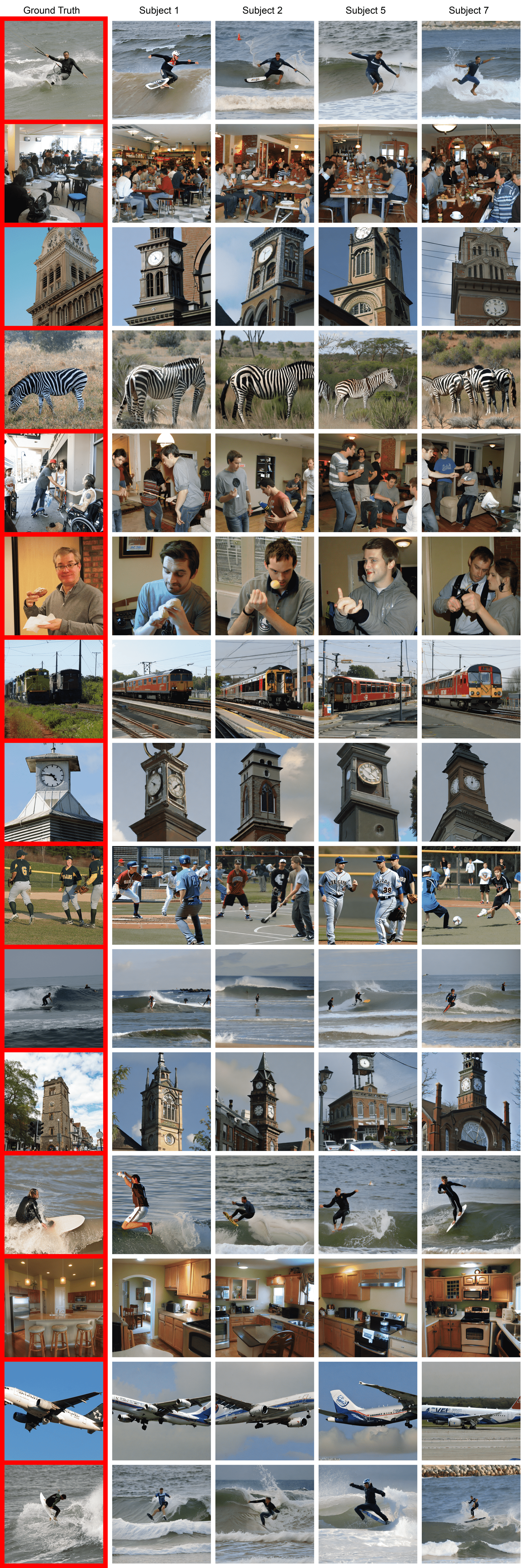}
    \caption{Some additional samples with comparison across subjects, ground truth images are indicated by a red background.}
    \label{fig:appendix2}
\end{figure}

\medskip
\newpage

\bibliographystyle{unsrt}

\bibliography{text}
}
\end{document}